# Electron Emission Yield Datasets Under Electron Impact From Surfaces Characterized In Situ by XPS or AES

M. Belhaj* and S. Dadouch**

DPHY, ONERA, Université de Toulouse, 31000, Toulouse, France

*Mohamed.Belhaj@onera.fr   **Sarah.Dadouch@onera.fr

## Abstract

We have generated at ONERA electron emission data under electron impact for several years, although only part of this work has been published. Existing tabulated datasets in the literature rarely document surface conditions, even though secondary electrons originate within only a few nanometers of the surface. Consequently, emission data cannot be reliably interpreted without detailed surface composition information, including contamination, oxidation after air exposure, or in situ cleaning. We present the measurement, characterization, and calibration procedures used to produce a series of datasets for various conductive and semiconductive materials. The data, provided, include emission yields as a function of incident electron energy together with surface composition obtained from X-Ray Photoelectron Spectroscopy (XPS) or Auger Electron Spectroscopy (AES) analyses. Initial datasets cover copper and gold, with additional materials (germanium, titanium, silver, aluminum, nickel, …) to be released on arXiv.

## INTRODUCTION

When exposed to electron bombardment, surfaces may in turn emit electrons that can be either secondary electrons, originating from the target material, or as backscattered electrons, corresponding to incident electrons that are re-emitted into the vacuum after undergoing sequences of elastic or inelastic interactions. The total number of emitted electrons may therefore be lower or higher than the number of incident electrons.

In many applications involving electron–surface interactions under vacuum, an accurate determination of the electron emission yield is essential. This yield is defined as the ratio between the total number of electrons emitted into the vacuum, including both secondary and backscattered electrons, and the number of incident electrons. It is denoted TEEY, Total Electron Emission Yield. The TEEY is the sum of the SEY, Secondary Electron Yield, and the BSEY, Backscattered Electron Yield. In the literature, especially in more recent publications, the TEEY is frequently referred to as SEY, a practice that unfortunately introduces confusion and errors.

Secondary electrons originate from depths on the order of a few nanometers. As a consequence, the SEY and thus the TEEY are extremely sensitive to the surface composition. Any contamination[1], oxidation[2], or intentional surface coating[3] results in a significant modification of the TEEY. In addition to chemical composition, the TEEY also depends on surface morphology[4], the angle of incidence[5], and, in certain cases, the temperature[6], as well as charge trapping phenomena in dielectric materials[7].

---

[1] Hilleret, N., Scheuerlein, C. & Taborelli, M. The secondary-electron yield of air-exposed metal surfaces. Appl Phys A 76, 1085–1091 (2003)

[2] Jiangtao Li, Bart Hoekstra, Zhen-Bin Wang, Jie Qiu and Yi-Kang Pu. Secondary electron emission influenced by oxidation on the aluminum surface: the roles of the chemisorbed oxygen and the oxide layer 2018 Plasma Sources Sci. Technol. 27 044002

The use of TEEY data extracted from the literature without detailed knowledge of the corresponding surface properties is therefore hazardous. It is essential to determine whether the surface was cleaned in situ or exposed to ambient air and therefore likely contaminated and possibly oxidized. It is also necessary to assess whether the surface was irradiated long enough during the TEEY measurement for the electron beam to alter its physicochemical properties. For dielectric materials, the injected and trapped charges, whether positive or negative, may also significantly influence the measured yield.

A comprehensive database, based on tabulated values of BSEY and SEY reported in the literature, was compiled by David D. Joy[8]. Although this database has been and continues to be extremely valuable to the scientific community, very limited information is available regarding the condition of the material surfaces, whether contaminated or not, produced under vacuum or not, eroded in situ, and so forth. Furthermore, the database clearly demonstrates, through the large discrepancies sometimes reaching 300 percent in the TEEY of a single material, the critical influence of surface properties.

These considerations illustrate the difficulty of relying on TEEY values reported in the literature. In many cases, answering the relevant questions is extremely difficult, if not impossible, due to the lack of information regarding the surface state and the measurement conditions typically accompanying published results.

The objective of this work is to provide a coherent set of TEEY data for several materials, or more precisely for specific surfaces, while supplying the maximum possible information concerning their surface characteristics.

## BASIS

Incident electrons interacting with a target undergo elastic scattering, which primarily redistributes their trajectories, and inelastic scattering, which induces processes such as collective excitations (plasmons) or individual electronic excitations. In dielectric materials, additional processes arise, including interactions with phonons and with trapped charges and their associated electric fields, which may even dominate under certain conditions.

Although electron excitations occur along the entire penetration depth of the incident electrons, which may extend up to approximately one micrometer for primary energies of several tens of keV, only those excitations produced within a few nanometers of the surface contribute to electron emission into the vacuum. To escape the material, electrons must overcome the potential barrier, defined by the work function for metals or the electron affinity for semiconductors and dielectric materials (+ the bandgap). The average energy of such excited electrons generally does not exceed a few tens of electronvolts. Once emitted into the vacuum, these electrons constitute the secondary electrons. Backscattered electrons, in contrast, may originate from deeper regions of the material[9].

---

[3] P. Costa Pinto et al, Carbon coatings with low secondary electron yield, Vacuum, Volume 98, 2013, Pages 29-36

[4] L. Diaz et al. Importance of surface morphology on secondary electron emission: a case study of Cu covered with carbon, carbon pairs, or graphitic-like layers. Sci Rep 13, 8260 (2023).

[5] N. Bundaleski et al Calculation of the angular dependence of the total electron yield, Vacuum, Volume 122, Part B, 2015, Pages 255-259,

[6] J. B. Johnson* and K. G. McKay, Secondary Electron Emission of Crystalline MgO, Phys. Rev. 91, 582 – Published 1 August, 1953

[7] Jacques Cazaux, About the secondary electron yield and the sign of charging of electron irradiated insulators, Eur. Phys. J. AP 15, 167-172 (2001)

[8] Lin, Yinghong and Joy, David. (2005). A new examination of secondary electron yield data. Surface and Interface Analysis. 37. 895 - 900. 10.1002/sia.2107.

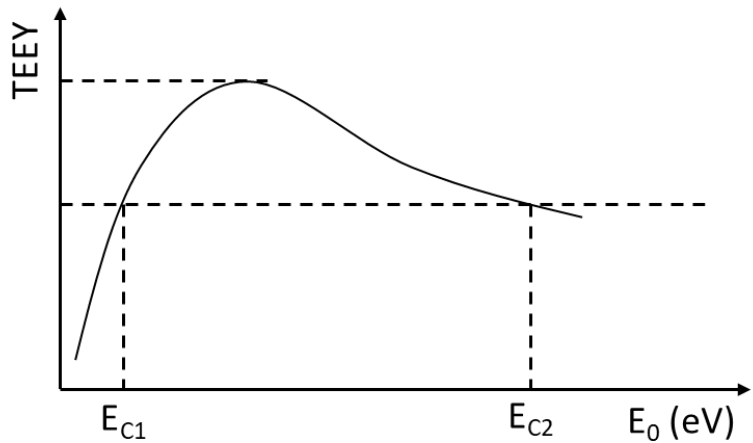

Figure 1- typical TEEY curve

A typical TEEY curve (Figure 1) as a function of the incident electron energy $E_0$ is characterized by three parameters:

• The maximum TEEY, $TEEY_{max}$, commonly used as an indicator of a material's propensity to emit electrons.

• The first crossover energy $E_{C1}$, when it exists, corresponding to the lowest incident energy at which TEEY=1.

• The second crossover energy $E_{C2}$, when it exists, corresponding to the higher energy at which TEEY=1.

Secondary electrons are predominantly low energy electrons, typically within a few eV, whereas backscattered electrons span a broad energy distribution extending up to $E_0$. An example of such a spectrum, measured on a copper sample, is provided in Figure 2.

---

[9] E. I. Rau et al, Electron backscattering coefficient from 3D nanostructures and determination of subsurface inhomogeneity depth in a scanning electron microscope. J. Appl. Phys. 28 October 2025; 138 (16): 165302.

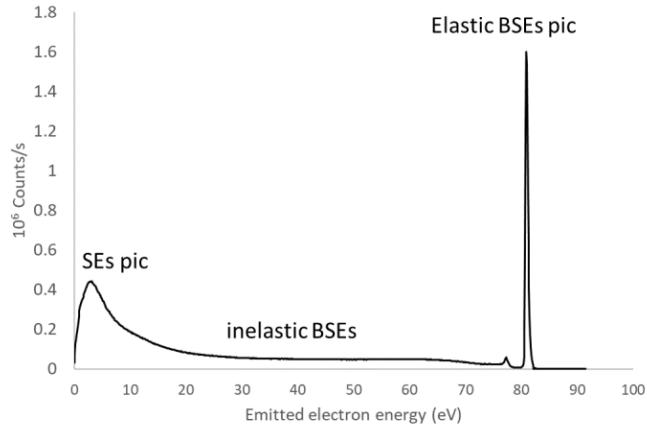

Figure 2- Energy distribution of the emitted electrons from the surface of Cu exposed to the atmosphere (ONERA/DEESSE data). The incident energy was 80 eV and the incidence angle were 45°.

# EXPERIMENTS AND METHODS

## Experimental Facilities

Two measurement systems, ALCHIMIE[10] and DEESSE[11], located at ONERA, were used. Both instruments operate under ultrahigh vacuum conditions ($10^{-10}$ to $10^{-9}$ mbar) and incorporate load-lock systems and analysis chambers enabling in situ TEEY measurements and surface analyses, including XPS, AES and REELS. They are equipped with electron guns covering the 1 eV to 30 keV range and ion guns covering 100 eV to 5 keV.

ALCHIMIE includes three Kimball Physics electron guns operating from 1 eV to 2 keV, from 1 keV to 30 keV, as well as a flood gun. DEESSE is equipped with a Kimball Physics gun (1 eV to 2 keV) and a STAIB gun (1 keV to 30 keV).

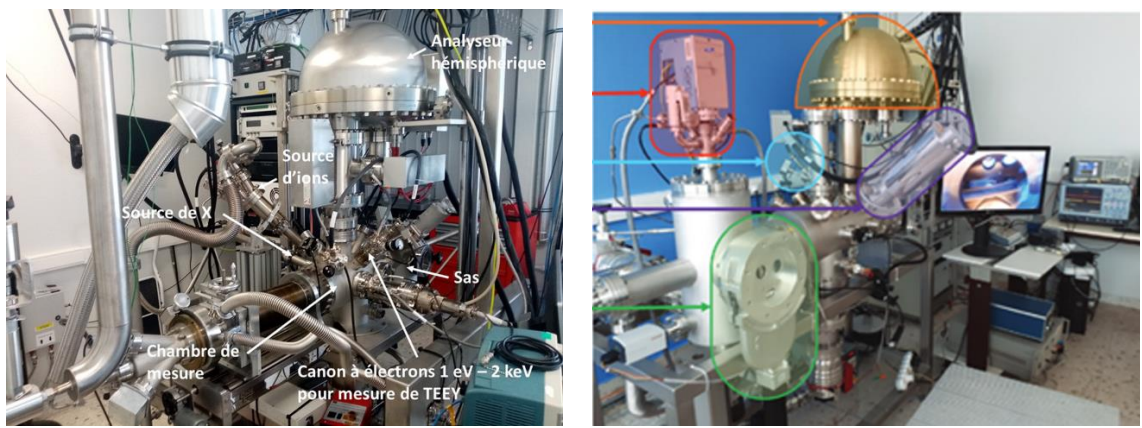

Figure 3-DEESSE and ALCHIMIE located at ONERA

---

[10] **A**na**L**yse **CH**Imique et **M**esure de l'é**M**Ision **E**lectronique
[11] **D**ispositif d'**E**tude de L'**E**mission **S**econdaire **S**ous **E**lectrons

Both systems allow variation of the incidence angle from 80° to normal incidence and offer temperature control of the sample holder, from −180 to +450°C for ALCHIMIE and from room temperature to +200°C degrees for DEESSE.

Each system includes a hemispherical electron analyzer, SIGMA for ALCHIMIE (128 MCP detector) and OMICRON EA125 (Cannelton detectors) for DEESSE, and both incorporate Faraday cups located on the sample holder and at the electron gun output.

The TEEY measurements and the surface characterizations (XPS or AES) were performed in the same analysis chamber without breaking the ultra-high vacuum.

TEEY Measurement Protocol

To avoid conditioning effects[12], that is electron-induced chemistry surface modifications, measurements are systematically performed in pulsed mode rather than continuous mode. For conductive materials, the pulse duration is on the order of one millisecond, whereas for dielectric materials it is on the order of one microsecond.

For each incident energy, ten pulses are recorded, from which the mean value and standard deviation are computed. If the standard deviation exceeds 3%, the measurement is discarded. The uncertainty increases as the incident energy decreases. The standard deviation as a function of incident energy is presented in Figure 4.

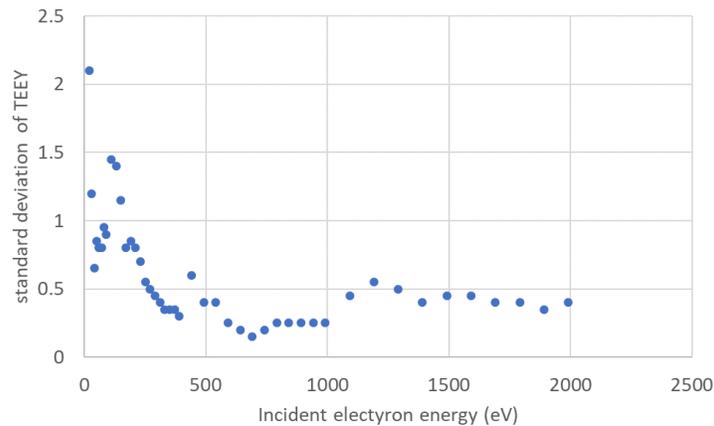

Figure 4- Typical standard deviation of TEEY as function of the incident electron energy

The measurement procedure consists of two steps. In the first step, the sample holder is positively biased. The sample is then bombarded with electrons. Secondary electrons with energies equal to or below +27 V, representing the majority of emitted electrons, are recollected by the surface. Time integration of the current measured on the sample holder provides the total charge $Q_0$. $Q_0$ is about few hundreds of pC per pulse for metals (pulse ms) and few hundreds of fC for dielectrics (pulse µs). The variation of Q0 as function of the incident electron energy is shown in Figure 5.

---

[12] V. Petit,*et al, Role of the different chemical components in the conditioning process of air exposed copper surfaces, Phys. Rev. Accel. Beams 22, 083101 (2019).

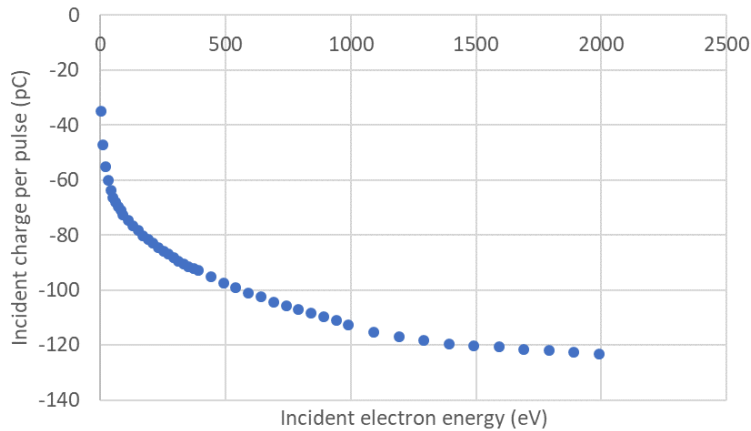

**Figure 5-Typical curve of the used Incident electron charge as function of the incident electron energy**

In the second step, the sample holder is negatively biased and the sample is again exposed to an electron flux with increasing primary energy. The negative bias ensures that no emitted electrons, especially secondary and tertiary[13] electrons, are recollected. The energy shift of the primary electrons, which are decelerated by the negative bias applied to the sample holder, is taken into account when plotting the TEEY curves.

The charge measured on the sample holder, $Q_S$, is related to the charge $Q_E$ associated with electron emission through $Q_E = Q_0 - Q_S$. Therefore,

TEEY = $1 - Q_S/Q_0$.

It should also be noted that measurements are generally performed with a constant focus setting of the electron gun. The focal point therefore varies as a function of the incidence energy. This implies that the analyzed area changes in a nonuniform manner with the incidence energy, typically from the millimeter to the centimeter scale.

XPS analyses Protocol

X ray radiation was generated using a non-monochromatic Al Kα line (1486.6 eV), with an Al anode polarization of 15 kV and an emission current of 20 mA.

The X ray source axis was set at 44° relative to the surface normal, and the electron energy analyzer angle was fixed at 10° with respect to the same reference.

Survey spectra were acquired in constant analyzer energy mode at 200 eV, with a typical number of passes ranging from 20 to 50.

High resolution spectra, when available, were collected in constant analyzer energy mode (20 eV), with a dwell time of 0.5 s, six passes, and a measurement step of 0.1 eV.

---

[13] Electrons generated when BSEs and SEs electrons interact with the inner surfaces of the analysis chamber.

## Comment on Dielectric Samples

Under electron irradiation, an insulator may acquire a positive charge when $E_{C1} < E_0 < E_{C2}$ or a negative charge when $E_0 < E_{C1}$ or $E_0 > E_{C2}$. The resulting surface charging is a major source of artefacts.

Positive charging accelerates the incident electrons and may lead to the recapture of secondary electrons. It also alters the transport of excited electrons within the material, reducing their mean free path due to interactions with accumulated charge and consequently decreasing their probability of emission[14].

Negative charging slows down or deflects incident electrons and similarly affects the transport of excited electrons. Moreover, some dielectrics may be pre-charged prior to measurement through triboelectrification.

These uncertainties, involving both the effective impact energy and the loss of secondary electrons, represent significant sources of artefacts.

The sample current method described above can be applied to insulating materials provided that appropriate adaptations are implemented. The electron dose delivered by the electron gun must be minimized, using the shortest possible pulses, typically of the order of one microsecond or less.

In addition to the total charge, the current density is critical, as broader beams reduce internal charging effects[15]. In the sample current method, the measured charge corresponds to the image charge induced by trapped charges rather than to a true flowing current. This requires specific experimental and geometrical conditions to ensure nearly complete electrostatic influence.

An alternative technique, known as the Kelvin Probe method[16], has recently been developed and improved. It consists of measuring the injected charge in situ using a surface potential probe and indirectly deducing the TEEY. This method provides a better evaluation of the injected charge and improved control of discharge processes[17].

Ongoing work aims to improve measurement techniques for dielectric materials. However, at present, we are not able to guarantee that TEEY measurements on dielectrics are entirely free of artefacts. We therefore choose not to release these data until the remaining uncertainties concerning their accuracy have been resolved.

---

[14] M. Belhaj et al, Effect of the incident electron fluence on the electron emission yield of polycrystalline Al2O3, Applied Surface Science, Volume 257, Issue 10, 2011, Pages 4593-4596
[15] Q. Gibaru et al ; Experimental and Monte-Carlo study of double-hump electron emission yield curves of SiO2 thin films. J. Appl. Phys. 7 April 2023; 133 (13): 135102
[16] M Belhaj et al 2009 J. Phys. D: Appl. Phys. 42 105309
[17] Alexandre Marcello Cavalca de Almeida et al 2025 J. Phys. D: Appl. Phys. 58 275205

# TEEY Calibration

## Absolute Value of the TEEY

Both instruments are regularly calibrated by comparing the TEEY measured at normal incidence on a 99.99 percent pure copper sample with the datasets reported by Petit et al[18]. and Cimeno et al[19]. Figure XXX shows the good overall agreement between these datasets. A slight deviation, up to eight percent, is observed between the three measurements for incident energies above 400 eV. This discrepancy may arise either from intrinsic measurement uncertainties or from differences

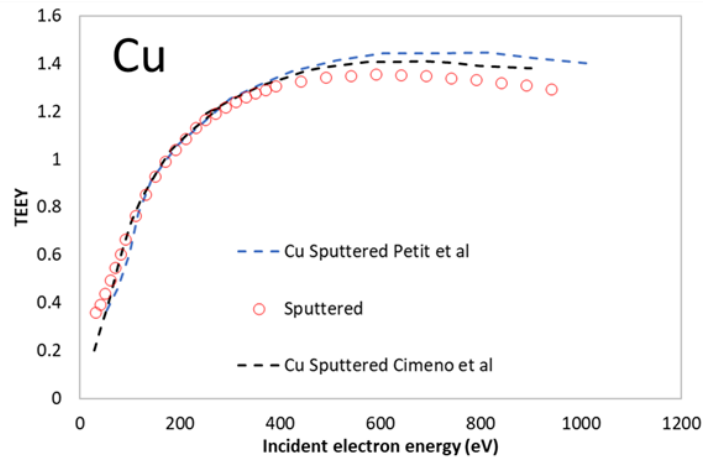

**Figure 6- Measured in DEESSE and ALCHIME TEEY curve of argon sputtered Cu compared to that measured in two others labs (INFN and CERN)**

## Incident Energy Calibration

The displayed beam energy may occasionally deviate from the actual impact energy, for example due to instabilities in the high voltage supply or other instrumental factors. Verification is therefore essential. When an electron energy analyzer is available in the analysis chamber, the energy of the elastic backscattered peak can be used to confirm the actual incident energy. Otherwise, the surface potential may be measured as described in earlier work[20]. Achieving an absolute energy accuracy of 1 eV is less straightforward than it may appear, since one must consider both the intrinsic energy dispersion of the electron gun, typically around 1 eV, and the difference between the work function of the sample and that of the gun cathode.

## Data

The term "*as received*" designates surfaces exposed to ambient atmosphere for an unspecified time at the supplier's, followed by several days to several months in our laboratory under ISO 8 conditions, without any subsequent cleaning. The term *etched* refers to surfaces cleaned using a 1 keV argon ion beam at normal incidence. Multiple etching steps may be carried out on the same sample. In such cases, intermediate TEEY, XPS and AES datasets are included in the accompanying Excel files.

---

[18] V. Petit et al, Role of the different chemical components in the conditioning process of air exposed copper surfaces Phys. Rev. Accel. Beams 22, 083101

[19] R. Cimeno et al Detailed investigation of the low energy secondary electron yield of technical Cu and its relevance for the LHC, Phys. Rev. ST Accel. Beams **18**, 051002

[20] M. Belhaj and S. Dadouch; A simple method for energy calibration of keV incident electron beam using a contactless electrostatic voltmeter probe. Rev. Sci. Instrum. 1 August 2021; 92 (8): 083301

Copper and gold samples were purchased from Goodfellow. They consist of 99.9% pure polycrystalline foils. Typical data are shown in the Figure 7.

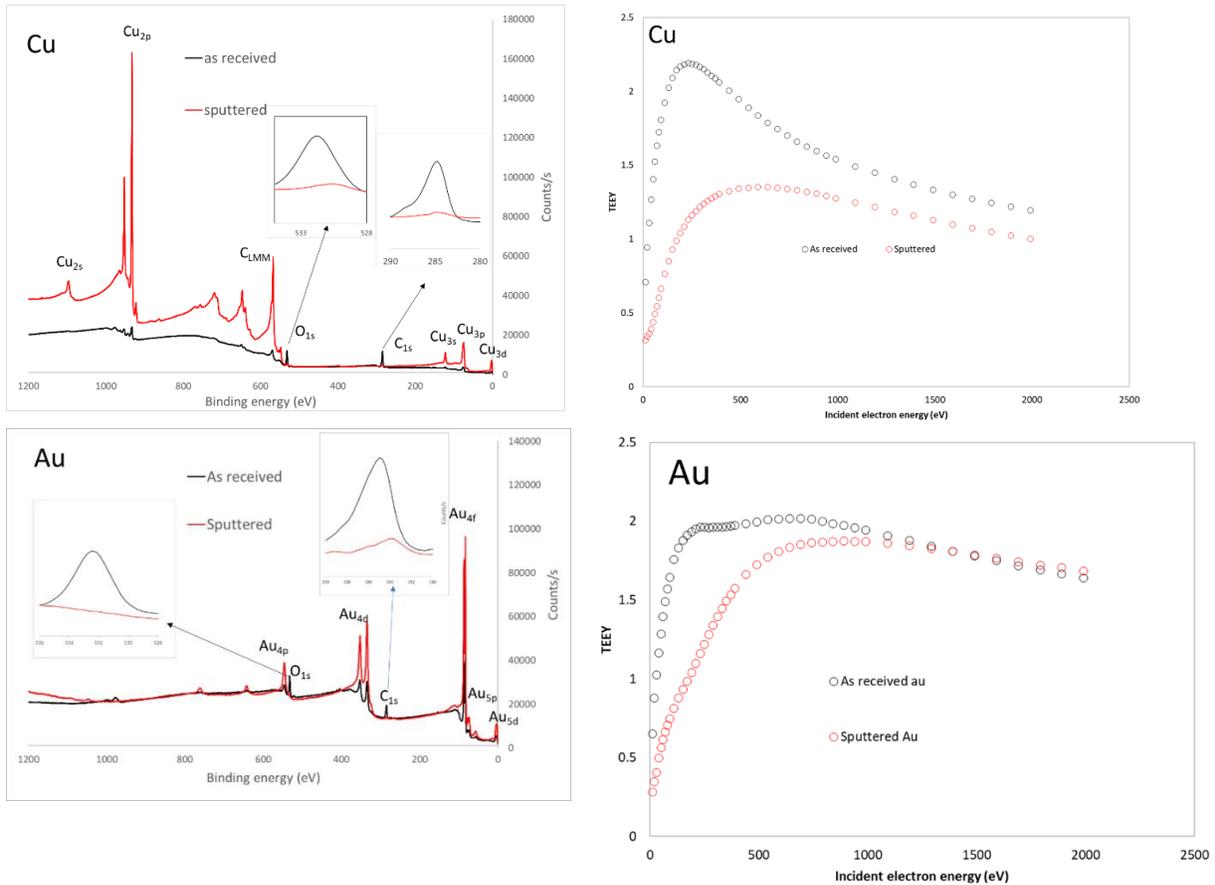

Figure 7- XPS spectra and TEEY of Cu and Au measured on the as received surface and etched surface

One of the xlsx files is shown in Figure 8. The first sheet contains two pairs of columns, the electron incidence energy and the TEEY, corresponding to the as-received surface and the eroded surface. The following sheets contain the XPS spectra for the as-received and eroded states, acquired in Survey mode and in high-resolution mode around the relevant spectral lines. Associated files in the CasaXPS (.vamas) format can be provided upon request.

Figure 8- XPS spectra and TEEY of Cu and Au measured on the as received surface and etched surface

## LICENSE AND TERMS OF USE

The datasets presented here are made available to the scientific community under the CC BY NC SA 4.0 license, in accordance with arXiv practices, to encourage their use whenever relevant.

Any use of these data in research outputs must include proper citation of the publication in which the dataset is originally described: M. Belhaj and S. Dadouch, Electron Emission Yield Datasets Under Electron Impact From Surfaces Characterized In Situ by XPS or AES, as well as to the associated DOI.